\documentclass[12pt]{article}

\usepackage{newtxtext,newtxmath}

\usepackage{graphicx}
\usepackage{xcolor}

\usepackage[letterpaper,margin=1in]{geometry}
\usepackage[utf8]{inputenc}
\DeclareUnicodeCharacter{1EC5}{$\tilde{\hat{\rm e}}$}

\renewenvironment{abstract}
	{\quotation}
	{\endquotation}

\date{}

\makeatletter
\renewcommand{\fnum@figure}{\textbf{Figure \thefigure}}
\renewcommand{\fnum@table}{\textbf{Table \thetable}}
\makeatother

\usepackage{scicite}

\usepackage{url}

\newcommand{\msun}{${\rm M}_\odot$}
\newcommand{\lsun}{${\rm L}_\odot$}
\newcommand{\kms}{km\,s$^{-1}$}
\newcommand{\degr}{$^{\circ}$}
\newcommand{\mas}{{\rm mas}}
\newcommand{\apj}{ApJ}
\newcommand{\apjl}{ApJL}     
\newcommand{\aap}{A\&A}
\newcommand{\mnras}{MNRAS}
\newcommand{\araa}{ARA\&A}
\newcommand{\ssr}{SSRv}
\newcommand{\pasp}{PASP}
\newcommand{\aj}{AJ}

\def\scititle{
	Massive extended streamers feed high-mass young stars
}
\title{\bfseries \boldmath \scititle}

\author{
    Fernando~A.~Olguin$^{1,2,3\ast}$,
    Patricio~Sanhueza$^{4}$,
    Adam~Ginsburg$^{5}$,\and
    Huei-Ru~Vivien~Chen$^{3}$,
    Kei~E.~I.~Tanaka$^{6}$,
    Xing~Lu$^{7}$,
    Kaho~Morii$^{8,9,2}$,\and
    Fumitaka~Nakamura$^{2,9}$,
    Shanghuo~Li$^{10,11,12}$,
    Yu~Cheng$^{2}$,
    Qizhou~Zhang$^{8}$,\and
    Qiuyi~Luo$^{7,13}$,
    Yoko~Oya$^{1}$,
    Takeshi~Sakai$^{14}$,
    Masao~Saito$^{2,15}$,
    Andr\'es~E.~Guzm\'an$^{16}$\and
    \small$^{1}$Center for Gravitational Physics, Yukawa Institute for Theoretical Physics, Kyoto University, \\
    \small Kitashirakawa Oiwakecho, Sakyo-ku, Kyoto 606-8502, Japan.\and
    \small$^{2}$National Astronomical Observatory of Japan, National Institutes of Natural Sciences,\\
    \small 2-21-1 Osawa, Mitaka, Tokyo 181-8588, Japan.\and
    \small$^{3}$Department of Physics and Institute of Astronomy, National Tsing Hua University, Hsinchu 30013, Taiwan.\and
    \small$^{4}$Department of Astronomy, School of Science, The University of Tokyo, 7-3-1, Hongo, Bunkyo-ku, \\
    \small Tokyo 113-0033, Japan\and
    \small$^{5}$Department of Astronomy, University of Florida,  P.O. Box 112055, Gainesville, FL 32611, USA.\and
    \small$^{6}$Department of Earth and Planetary Sciences, Institute of Science Tokyo, Meguro, Tokyo 152-8551, Japan.\and
    \small$^{7}$Shanghai Astronomical Observatory, Chinese Academy of Sciences, 80 Nandan Road, Shanghai 200030,\\
    \small People’s Republic of China.\and
    \small$^{8}$Center for Astrophysics $|$ Harvard \& Smithsonian, 60 Garden Street, Cambridge, MA 02138, USA.\and
    \small$^{9}$Department of Astronomy, Graduate School of Science, The University of Tokyo, 7-3-1 Hongo, Bunkyo-ku,\\
    \small Tokyo 113-0033, Japan.\and
    \small$^{10}$School of Astronomy and Space Science, Nanjing University, 163 Xianlin Avenue, Nanjing 210023, China.\and
    \small$^{11}$Max Planck Institute for Astronomy, Königstuhl 17, D-69117 Heidelberg, Germany.\and
    \small$^{12}$Key Laboratory of Modern Astronomy and Astrophysics (Nanjing University), Ministry of Education,\\
    \small Nanjing 210023, People’s Republic of China.\and
    \small$^{13}$School of Astronomy and Space Sciences, University of Chinese Academy of Sciences,\\
    \small No. 19A Yuquan Road, Beijing 100049, P. R. China.\and
    \small$^{14}$Graduate School of Informatics and Engineering, The University of Electro-Communications,\\
    \small Chofu, Tokyo 182-8585, Japan.\and
    \small$^{15}$Astronomical Science Program, The Graduate University for Advanced Studies, SOKENDAI,\\
    \small 2-21-1 Osawa, Mitaka, Tokyo 181-8588, Japan.\and
    \small$^{16}$Joint Alma Observatory (JAO), Alonso de Córdova 3107, Vitacura, Santiago, Chile.\and
	\small$^\ast$Corresponding author. Email: f.olguin@yukawa.kyoto-u.ac.jp
}

\begin{document} 

\maketitle

\begin{abstract} \bfseries \boldmath
Stars are born in a variety of environments that determine how they gather gas to achieve their final masses.
It is generally believed that disks are ubiquitous around protostars as a result of angular momentum conservation and are natural places to grow planets.
As such, they are proposed to be the last link in the inflow chain from the molecular cloud to the star.
However, disks are not the only form that inflows can take.
Here we report on high-resolution observations performed with the Atacama Large Millimeter/submillimeter Array that reveal inflows in the form of streamers.
These streamers persist well within the expected disk radius, indicating that they play a substitute role channeling material from the envelope directly to an unresolved small disk or even directly to the forming high-mass protostar.
These flows are massive enough to feed the central unresolved region at a rate sufficient to quench the feedback effects of the young massive star.
\end{abstract}


\subsection*{Teaser}
\noindent Infalling streamers may play a role as critical as disks in delivering gas to fuel the growth of young high-mass stars.

\subsection*{MAIN TEXT}

\subsection*{Introduction}

Stars form in the densest regions of turbulent and magnetically threaded molecular clouds, namely cores.
Typical cores have sizes of $2000-4000$\,au \cite{Motte2018,Motte2022,Louvet2024,Ishihara2024} and consist of an envelope that feeds a central disk \cite{Beltran2022} formed as a result of the conservation of angular momentum.
Close to the center, a stellar system eventually forms, fed by gas flowing through these structures.
One important consequence of the interplay among the magnetic field, turbulence, and self-gravity is the development of anisotropic infall in the form of streams that can promote the transport of gas to the disk and subsequently to the star(s) \cite{Seifried2015,Commercon2024}. 
Infall streams (hereafter streamers) have been observed across several star formation mass and spatial regimes \cite{Pineda2020,Karnath2020,Sanhueza2021,Sanhueza2025,Fernandez-Lopez2023,ValdiviaMena2024,Morii2025}, yet their role is still under debate \cite{Pineda2023,Tobin2024}.
In addition to the anisotropic infall described above, other paths for the formation of streamers include cloudlet capture \cite{Dullemond2019,Hanawa2022} and star-cloud interactions \cite{Yano2024}.

In the high-mass regime, the high luminosity hinders or even reverses spherically symmetric infall, and nonspherical accretion is needed to allow the star to grow by letting the stellar radiation and winds to escape through the polar directions \cite{Yorke2002,Tanaka2017,Kuiper2018,Commercon2022,Rosen2022}.
Here we show the case of the Galactic high-mass core G336.018-00.827 (hereafter G336 ALMA1; distance 3.1\,kpc \cite{Urquhart2018}) where non-spherical accretion seems to be attained predominantly from streamers without the presence of a large Keplerian disk as those detected in other regions (e.g., \cite{Rosen2020,Ohashi2023}).
Previous Atacama Large millimeter/submillimeter Array (ALMA) 1.3\,mm observations at ${\sim}0.05$'' resolution revealed the presence of three continuum peaks, two of which are connected to streamers.
These streamers bring gas from 2000\,au scales down to a radius of 400\,au, which corresponds to the radius at which the centrifugal and gravitational forces balance each other (hereafter centrifugal radius),  at a rate of roughly $10^{-4}$\,\msun\,yr$^{-1}$ \cite{Olguin2023}. 
A central mass of 10\,\msun\ was determined by \cite{Olguin2023} from the modeling of the gas kinematics under an infalling and rotating envelope (IRE; \cite{Oya2022}).
Those observations showed that streamers can account for a substantial part of the total infall rate from the large-scale envelope down to few 100\,au scales.

\subsection*{Results and Discussion}

\subsubsection*{Higher-resolution ALMA observations}

\begin{figure}[hp]
\centering
\includegraphics[width=\textwidth]{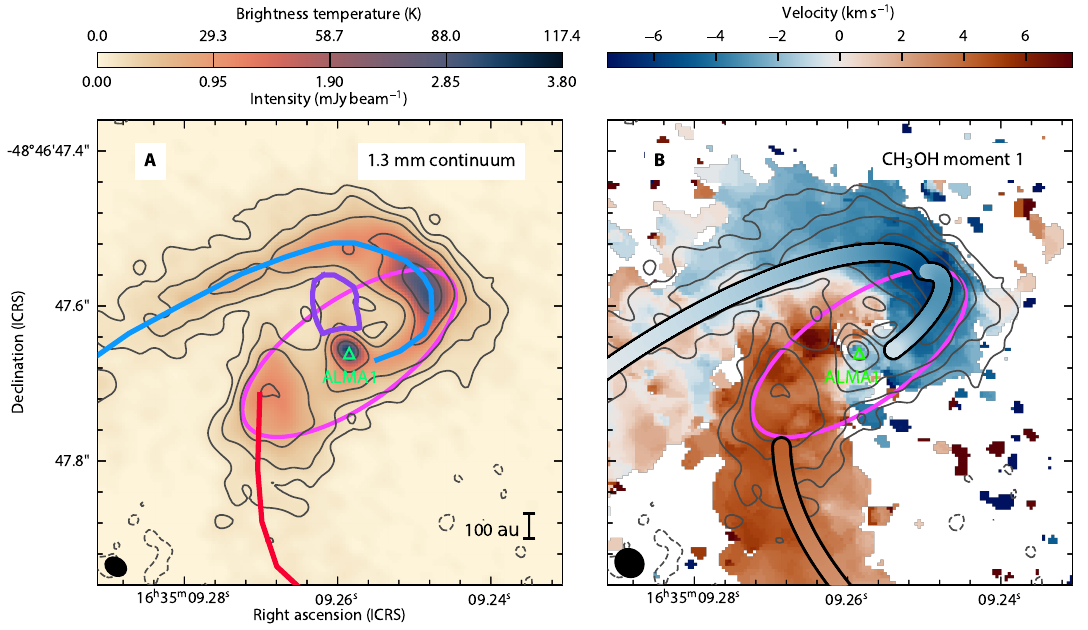}
\caption{\textbf{ALMA 1.3\,mm continuum emission and CH$\bf _3$OH $\bf J_{K_a,K_c} =18_{3,15}-17_{4,14}\,A,\,v_t=0$ first moment map.} 
\textbf{(A)} The green triangle corresponds to the central source, ALMA1.
The contour levels are $-3$, 5, 10, 20, 40, 80$\times\sigma_{rms}$ with $\sigma_{rms}=28$\,$\mu$Jy\,beam$^{-1}$.
The blue- and red-shifted streamer directions are shown by the curves for the corresponding color.
The purple region shows extended emission from ALMA1 likely tracing a jet component.
\textbf{(B)} The first moment velocity map with respect to the systemic velocity ($-47.2$\,km\,s$^{-1}$ \cite{Taniguchi2023}) is accompanied by colored curves describing the trajectory and velocity distribution of the streamlines of a rotating and infalling envelope.
The blue streamer is divided in two streamlines.
The outer component presents a refinement of the streamline model presented in \cite{Olguin2023}, while the inner component corresponds to a similar model but with a streamline origin at the centrifugal radius and close to the mid-plane (see Materials and Methods).
The contours are the same as in (A).
The pink ellipse correspond to a circle of radius 500\,au, i.e., the centrifugal radius, projected in plane of the sky assuming an inclination angle of 65\degr\ with respect to the line of sight.
Synthesized beam sizes for the continuum and first moment maps are shown in the bottom left corner.
}\label{fig:continuum}
\end{figure}

To investigate the small-scale structure of this region and to search for the presence of an accretion disk, we conducted high-resolution (28\,mas or 86\,au, Fig.~\ref{fig:continuum}A; see Material and Methods) observations of the 1.3 mm continuum and several molecular lines.
The system consists of a single central source (hereafter ALMA1) of deconvolved size (FWHM)  39\,mas (121\,au; see Materials and Methods), connected to the blue-shifted streamer to the west.
To the east no clear connection to the red-shifted streamer is observed in the continuum, only faint emission is revealed.
Extended continuum emission is observed to the north of the source over the 10$\sigma$ detection level (purple region in Fig.~\ref{fig:continuum}A), which may be associated with free-free emission of a jet (see below).
While we detect continuum emission from the three density peaks and the northern streamer previously seen at 50\,mas resolution, we detect no conventional disk at the centrifugal radius from the central peak.  
Here we define a conventional disk as an over-dense structure of projected elliptical shape with a Keplerian velocity gradient due to rotation, such as those found toward low-mass sources from the eDsik collaboration \cite{Ohashi2023} or toward high-mass sources like GGD 27-MM1 \cite{Fernandez-Lopez2023} and G11.92 \cite{Ilee2018}.
Although a velocity gradient does indicate rotation around the central source (see below), this gradient is found only in the gas associated with the streamers, and not in a flattened, symmetric disk structure.
The unresolved central source could contain a disk or another over-dense structure (e.g., a torus).
Disk candidates with similar small sizes and extended spiral-like emission were identified in W51 (radii $<500$\,au with W51e8 smaller than 75\,au; \cite{Goddi2020}), but no infall signatures were detected along the spiral structures.
Other high-mass sources (e.g., AFGL 4176 mm1 \cite{Johnston2020}, Sgr C \cite{Lu2022}) have also shown substructures similar to spiral arms within a disk (${\sim}1000$\,au radius in the case of AFGL 4176 mm1 and Sgr C), and the continuum emission of the low-density disk can be identified. 
However, this is not the case in the continuum of G336 ALMA1.

\subsubsection*{Rotating and infalling motions}

\begin{figure}[hp]
\centering
\includegraphics[width=\textwidth]{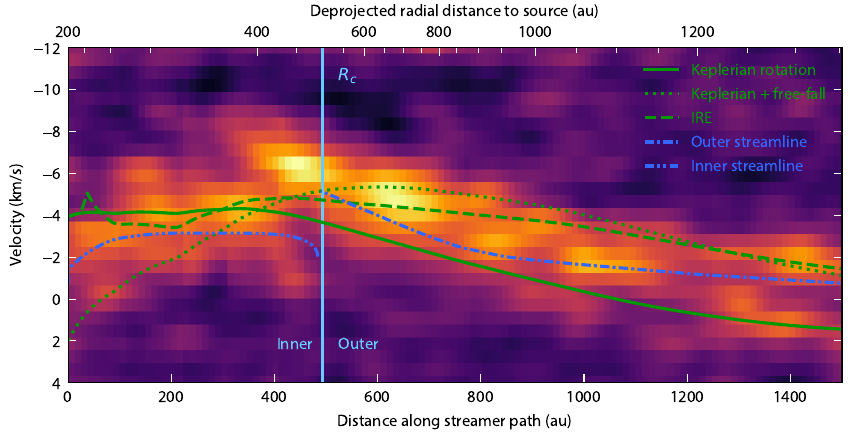}
\caption{\textbf{Blue-shifted streamer position-velocity (PV) diagram in CH$\bf _3$OH.}
The green solid line corresponds to the line of sight Keplerian velocity at the same deprojected distance to the central source, ALMA1.
Similarly, the dotted green line corresponds to a Keplerian rotation and infall velocity distribution, while the dashed line corresponds to the IRE velocity distribution \cite{Oya2022}.
The inner and outer blue streamer models in Fig.~\ref{fig:continuum}B are shown in blue dot-dashed lines.
The lower abscissa scale corresponds to the distance along the path used to calculate the PV diagram slice (shown in Fig.~\ref{fig:continuum}) with the package  \textsc{pvextractor}, while the upper one corresponds to the radial distance to ALMA1 assuming that the slice points are located in the mid-plane (see Materials and Methods).
The mid-plane is assumed to be inclined 65\degr\ with respect to the line of sight.
The vertical light blue line marks the position of the centrifugal radius, $R_c=500$\,au.
}\label{fig:pvmap}
\end{figure}

\begin{figure}[tb]
\centering
\includegraphics[width=\textwidth]{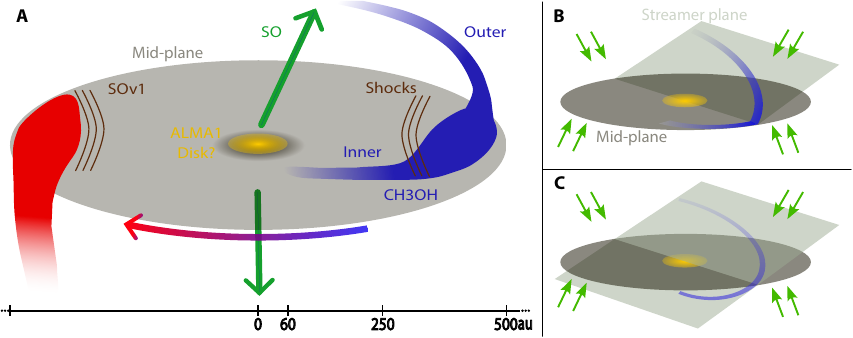}
\caption{\textbf{Schematic representation of the different kinematic components and flow scenarios.}
({\bf A}) Schematic representation of the observations with the different components labeled by the tag with the respective color.
The green arrows point in the direction of the outflows, while the blue to red arrow indicated the rotation direction.
The scale at the bottom indicates the radius of ALMA1, the centrifugal barrier radius and the centrifugal radius from left to right, respectively.
({\bf B}) and ({\bf C}) Are the two scenarios proposed to explain the origin of the shocks in the blue streamer.
Green arrows represent the background infall from the less dense envelope.
}\label{fig:sketch}
\end{figure}

In addition to the continuum morphology, molecular line emission from hot methanol (CH$_3$OH $J_{K_a,K_c} =18_{3,15}-17_{4,14}\,A,\,v_t=0$, $E_{u}/k_{\rm B}=447$\,K; Fig.~\ref{fig:continuum}B) also shows a clear connection between the blue-shifted streamer and the central young star ALMA1.
Like the continuum emission, the methanol emission shows no evidence for a large disk. (see also fig.~\ref{fig:supp:mom0}).
This emission reveals a change in the velocity gradient roughly at the position of the centrifugal radius (updated to 500\,au from 400\,au \cite{Olguin2023}, see Materials and Methods) which could result from a change in the flow direction and/or in the velocity distribution (see also Fig.~\ref{fig:pvmap}).
In the following analysis, we will refer to  the blue-shifted streamer sections within and outside a radius of 500\,au from ALMA1 as the inner and outer blue streamers, respectively. 
The different components of the system are summarized in the sketch in Fig.~\ref{fig:sketch}A.
While methanol emission from the red-shifted streamer may indicate a connection with ALMA1, we exclude it because of contamination from the outflow.

To understand what is happening in the western (blue-shifted)  inner streamer, we model the infall from the centrifugal radius inward with the same rotating and infalling streamlines equations of \cite{Mendoza2009} (see also \cite{Pineda2020} for their implementation).
The model describes relatively well the trajectory of the inner blue streamer (Fig.~\ref{fig:continuum}B) assuming it is in the mid-plane defined by the rotation axis (Fig.~\ref{fig:sketch}A) with an initial radial velocity close to zero (0.1\,\kms; see Materials and Methods).
However, the velocity profile does not match, as it underestimates the velocity along the line of sight.
We compare the observed position-velocity (PV) diagram (Fig.~\ref{fig:pvmap}) along the spine of the blue-shifted streamer (shown in blue in Fig.~\ref{fig:continuum}A) with the streamline models (inner and outer) and three other velocity distributions (see equations in Materials and Methods): Keplerian rotation, Keplerian rotation with  free-fall infall, and the velocity distribution of an IRE \cite{Oya2022}. 
These distributions are evaluated at the same deprojected radial distance to ALMA1 assuming an inclination angle of 65\degr\ (\cite{Olguin2023}; green lines in Fig.~\ref{fig:pvmap}).
We find that Keplerian rotation and IRE are the best-fitting models for the inner blue streamer velocity, with Keplerian rotation providing a better fit to the peaks at each offset value.
A Keplerian velocity is expected after the formation of a disk, once the infalling gas has settled (e.g., \cite{Oliva2020}).
In the case of the IRE model, the radial distances between ALMA1 and the spine of the inner blue streamer, i.e., between the centrifugal radius and the centrifugal barrier (the radius at which the kinetic energy is converted into rotational energy, and equal to half the centrifugal radius), correspond to a transition zone where the relative contribution of infall decreases and rotation increases as the radius decreases.
Thus the observed velocities are likely dominated by a rotational component as expected for a disk-like structure.
For the outer blue streamer, Keplerian rotation alone cannot explain the observed velocity distribution, which can only be explained by the contribution of the infall as shown by the other velocity distributions as well as by the streamer modeling  (Fig.~\ref{fig:continuum}B and \ref{fig:pvmap}).

\subsubsection*{Shocked gas}\label{sect:shocked}

\begin{figure}[tb]
\centering
\includegraphics[width=\textwidth]{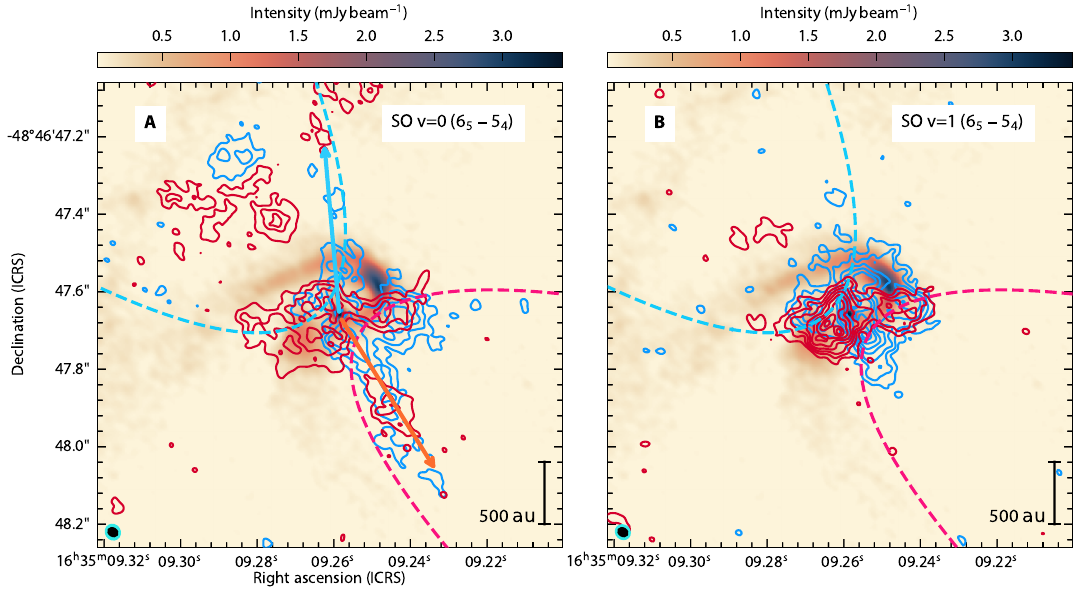}
\caption{\textbf{Blue- and red-shifted SO emission in contours over 1.33\,mm continuum map.}
Blue and red contour levels are 3, 4, 5, ...$\times\sigma$ with a $\sigma=4.8$\,mJy\,beam$^{-1}$\,km\,s$^{-1}$ from the data with worst noise (SO $v{=}1$). 
The directions and approximate position of the northern and southern outflow cavities are traced by the blue and red dashed parabolas, respectively.
The blue and orange arrows indicate the direction of shocked emission likely associated with a jet component.
For a more detailed view of the jet/outflow components see fig.\,\ref{fig:supp:sov0}.
The synthesized beams of the continuum (black) and SO contours (teal) are shown in the bottom left corner.
}\label{fig:so}
\end{figure}

Along the streamer, the position of the projected velocity change (Fig.~\ref{fig:pvmap}) roughly coincides with the location of the western continuum peak (see Figs.~\ref{fig:continuum} and \ref{fig:supp:model}). 
On that location, the emission could be enhanced by an increase in the column density, as implied by the methanol line emission, but also by an increase in temperature due to shocks.
To explore shocked gas, we inspect the SO $^3\Sigma\, J_K=6_5-5_4$ emission in its ground ($E_u/k_{\rm B}=35$\,K) and vibrationally ($E_u/k_{\rm B}=1634$\,K) excited states (see Materials and Methods).
Fig.~\ref{fig:so} shows blue- and red-shifted emission around the systemic velocity.
We find that in the ground state SO traces mostly gas in the outflow, in particular it traces the base of a parabolic bipolar outflow cavity (see also Figs.~\ref{fig:supp:sov0} and \ref{fig:supp:sov1}).
Additionally, extended blue-shifted emission  is observed to the north of ALMA1, this coincides with the continuum emission in that direction (marked in purple in Fig.~\ref{fig:continuum}A).
Hence that continuum emission may be associated with a jet rather than another feature located in the mid-plane or envelope.
Note that blue- and red-shifted emission is detected towards both outflow directions, probably indicative of a rotating outflow.
Based on the morphology of the inner blue streamer, the viewing angle is through the northern pole of the outflow system, contrary to what was assumed based on SiO outflow emission in \cite{Olguin2023}.

\begin{figure}[tb]
\centering
\includegraphics[width=\textwidth]{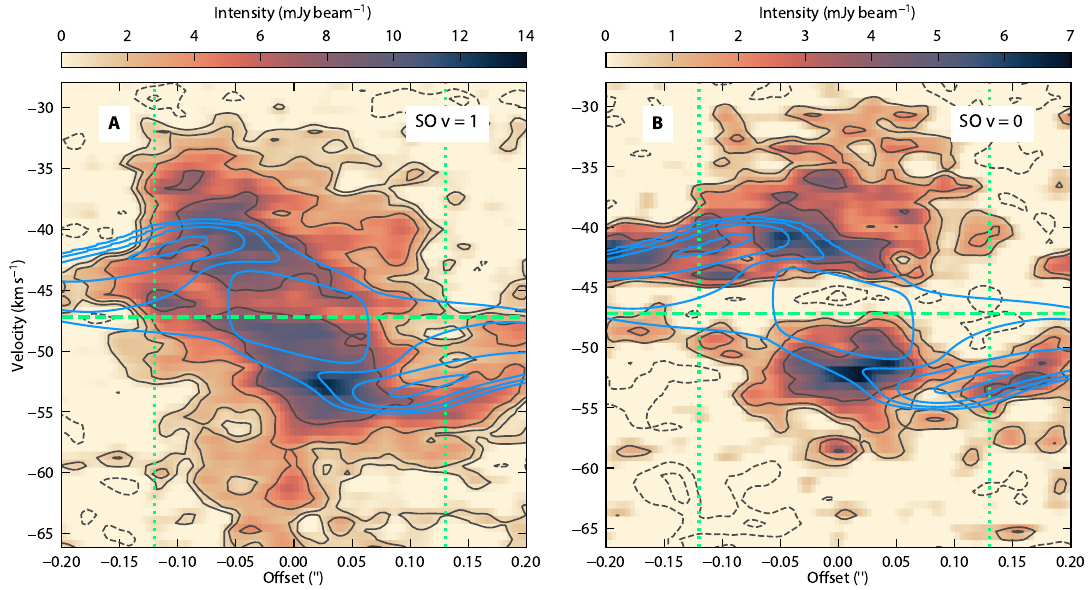}
\caption{\textbf{SO $\bf ^3\Sigma$ position velocity maps and IRE model.} 
The gray contour levels of SO emission are $-10$, $-5$, 3, 6, 12, 24$\times\sigma$ with $\sigma=0.3$\,mJy\,beam$^{-1}$.
The blue contours correspond to the IRE model and are at levels 3, 6, 12 and $24\times\sigma$ with $\sigma=1.2$\,mJy\,beam$^{-1}$ to match the CH$_3$OH data (see fig.~\ref{fig:supp:model}).
The horizontal green dashed line corresponds to the systemic velocity ($-47.2$\,\kms), and the vertical dotted lines correspond to the offsets of the continuum peaks ($-0.12$'' for the red-shifted streamer and 0.13'' for the blue-shifted one).
\label{fig:pvmap:so}}
\end{figure}

The vibrationally excited SO emission is more compact than its ground state (Fig.~\ref{fig:so}), and seems to be tracing outflow and shocks around the centrifugal radius components.
Contrary to the ground state, emission is observed towards the western continuum peak.
Along the envelope position angle (${\rm P.A.}=-55$\degr; see Materials and Methods), the PV diagram (Fig.~\ref{fig:pvmap:so}A) shows that the emission peaks at an offset of roughly 0.05'' (155\,au), which in turn corresponds to roughly half the offset of the continuum peaks, and around the centrifugal barrier. 
This is inside the position where the blue- and red-shifted streamers join the mid-plane (centrifugal radius).
This may be indicative of shocks as the gas from the streamer joins the mid-plane, and in turn explains the accumulation of gas at those locations (resulting in higher column densities and temperatures).
To explain the nature of the shocks, we propose two alternatives summarized in Fig.~\ref{fig:sketch}B and C.
Both imply that the shock is produced by the interaction between the streamer and the background infalling gas (from the less dense envelope component) as the streamer reaches the mid-plane.
On scenario Fig.~\ref{fig:sketch}B, the shock modifies the gas velocity and flow direction, while in Fig.~\ref{fig:sketch}C the gas crosses the mid-plane and continues in a spiral trajectory with the change in velocity explained solely by the projected trajectory along the line of sight.

\subsubsection*{Gathering gas and feeding high-mass protostars}

In this section we estimate the mass accretion rates and time scales for building up the observed mass of the central star, and show that the flow is sufficient to overcome radiative feedback and allow the star to continue to grow.
The case of G336 ALMA1 shows that streamers can play an important role in feeding high-mass protostars.
The continuity of the flow inward from ${\sim}$2000 to 60\,au (albeit with a change in velocity distribution) may indicate that the inner blue streamer is not a feature induced by the outer blue streamer in the gas already in the mid-plane or in a former disk, but rather a continuation of that flow.
The already large mass attained by the young star (10\,\msun\ \cite{Olguin2023}) implies that radiation feedback may have already begun.
In order to continue accreting gas, the density around the source has to be high enough to quench the feedback from the young star or the momentum carried by the streamers be high enough to overcome the feedback in the absence of a disk (e.g., \cite{Rosen2022}).
Considering that accretion is still taking place, as implied by the presence of jets/outflows, we estimate an upper-limit for the accretion rate from the free-fall time and lower-limit under viscous accretion. 
The time it would take to accrete the mass in the potential 60\,au disk around the young star (ALMA1) is in the ${\sim}25-10^4$\,yr range (see Materials and Methods). 
Similarly, we estimate the amount of time it would take the streamers to achieve the same central mass.
For a free-falling flow, it would take the streamers ${\sim}150$\,yr to replenish the gas in ALMA1. 
This is one order of magnitude longer than the free-fall time of ALMA1 (${\sim}25$\,yr).
On the other hand, for a viscous flow the replenishing time and accretion times are of the same order of magnitude ($10^3-10^4$\,yr) for the same viscosity parameter value.
Given that the infall is likely neither free-falling nor viscous, the real values should be in between our estimates.
These results indicate that the gas around the star in ALMA1 could be depleted faster than the streamers can replenish it in the free-fall case scenario.
In order to demonstrate that even in this scenario the streamers can overcome the stellar feedback, we calculate the magnitude of the forces exerted by the radiation field of the central young star and that of the blue-shifted streamer inflow.
We obtain that the blue-shifted streamer can exert a force two orders of magnitude higher than that of the radiation field red (see Materials and Methods).
Thus, the blue-shifted streamer can likely overcome the radiation pressure (at a radius of 61\,au) and eventually feed the young star.
In addition, the red-shifted streamer, though less massive, can still provide gas at a rate roughly half that of the blue-shifted streamer.

We estimate masses between 0.3--0.6\,\msun\ for each inner streamer (see Materials and Methods).
These masses and the resulting infall rates (${\sim}10^{-3}$\,\msun\,yr$^{-1}$) are an order of magnitude or higher than those found in streamers feeding low-mass star (e.g.,  \cite{ValdiviaMena2022,Hsieh2023,Thieme2022}).
G336 ALMA1 is located at the center of a massive clump of 350\,\msun\ (\cite{Csengeri2017}, previously classified as a core).
While a small (Keplerian) disk similar to those inferred in W51 \cite{Goddi2020} and in G345 \cite{Guzman2020} can still be the last link to feed the star, its mass is comparable to or lower than that of the streamers. 
It is thus the large mass of the reservoir, at large scales, and the streamers, at small scales, that have allowed the formation and continuous feeding of the young high-mass star at the center of ALMA1.

\subsection*{Materials and Methods}\label{sect:methods}

\subsubsection*{Observations and data reduction}

G336.018-0.827 was observed with the 12m Atacama Large millimeter/submillimeter Array  (ALMA) as part of the Digging into High-mass Cores with ALMA (DIHCA) project on July 2019 (ID: 2017.1.00237.S) and as part of follow-up observations on July 2023 (ID: 2022.1.00700.S).
Detailed description of the former data set are described in \cite{Olguin2023}, suffice to say the observations were performed in configuration C-8 with 45 antennas.
The most recent observations were performed with the C-9 array configuration with 49 antennas.
The delivered calibrated data was combined with the DIHCA observations and self-calibration was performed on the combined data set.
Only phase self-calibration with increasingly smaller solution intervals were used to produce the final calibrated data.

Both observations were made with a channel width of 488.281\,kHz, which corresponds to a spectral resolution (2 channels) of ${\sim}1.5$\,km\,s$^{-1}$, but their spectral coverage are slightly shifted.
Hence, the spectral windows were trimmed down to the overlapping ranges and the channels aligned using the CASA \cite{2022PASP..134k4501C} task CVEL.
We then followed the procedure described in \cite{Olguin2021} (\textsc{GoContinuum} \cite{fernando_2025_15314190}) to obtain line-free visibilities and continuum subtracted visibilities.
The continuum was imaged using the TCLEAN task with the Hogbom deconvolver, auto-masking and a Briggs weighting with robust parameter of 0.5.
The continuum is cleaned down to a threshold of $1.25\sigma$ with the noise $\sigma$ determined by TCLEAN.
The resulting continuum angular resolution is $0.031\times0.025$'' (P.A.=57.5\degr; $96\times78$\,au) and noise level is $28\,\mu$Jy\,beam$^{-1}$.

In order to optimize the computing resources, we only imaged selected molecular lines with the auto-masking routine YCLEAN \cite{Contreras18,2018zndo...1216881C}: CH$_3$OH $J_{K_a,K_c} =18_{3,15}-17_{4,14}\,A,\,v_t=0$ (233.795666\,GHz, $E_u/k_{\rm B}=447$\,K), SO $^3\Sigma\,v=0\,J_K=6_5-5_4$ (219.949442\,GHz, $E_u/k_{\rm B}=35$\,K) and SO $^3\Sigma\,v=1\,J_K=6_5-5_4$ (218.3238577\,GHz, $E_u/k_{\rm B}=1634$\,K). 
Given the larger spatial extent of some of these lines, we use the multi-scale deconvolver with scales 0, 5, 15 and 25, and Briggs weighting with robust parameter of 2 to increase the S/N of the lines.
The median resolutions of the cubes are $0.041\times0.037$'' (P.A.=47\degr; $127\times115$\,au) for CH$_3$OH, $0.045\times0.041$'' (P.A=32\degr; $140\times127$\,au) for SO $v=0$, and $0.044\times0.040$'' (P.A.=47\degr; $136\times124$\,au) for SO $v=1$.
Their noise levels are between 1.1--1.3\,mJy\,beam$^{-1}$ per channel.

\subsubsection*{Modeling refinement}

To set the central source position, we first fit a Gaussian to the ALMA1 continuum emission within a region of radius of 50\,mas with the CASA task IMFIT.
The central source right ascension is $16^{\rm h}35^{\rm m}09.2585^{\rm s}\pm0.0002^{\rm s}$ and its declination is $-48^{\circ}46'77.661''\pm0.002''$ from the fit.
The region has a deconvolved size (FWHM) of $43.5\pm8.6\,\mas\times35.1\pm8.6\,\mas$ (P.A.=28\degr), resulting in a geometric mean size of $39\pm6\,\mas$ (121\,au).
The flux density of the Gaussian is $7.7\pm1.1$\,mJy.

The higher angular resolution of the most recent observations allows us to refine the parameters of the models describing the rotation and infall along the streamers presented in \cite{Olguin2023} (outer component in Fig.~\ref{fig:sketch}A).
In particular, the observations allow us to pinpoint the position of the centrifugal radius and centrifugal barrier.
We use the streamline model for an infalling and rotating envelope described in \cite{Mendoza2009} and implemented by \cite{Pineda2020} with the same parameters as \cite{Olguin2023} but varying the centrifugal radius, initial polar angle, and initial azimuthal angle for the outer blue streamer.
Parameter ranges are summarized in Table~\ref{tab:streamline:params}.
It is worth noting that all the angles are in the \texttt{velocity\_tools} package standard, which can be differ with other implementations of the \cite{Mendoza2009} equations.
In this standard, the outflow position angle required to match the rotation direction is 210\degr\ and the inclination angle is 25\degr\ (defined as 0\degr\ for the edge-on configuration).
\cite{Olguin2023} found that centrifugal radii between 400--600\,au can fit relatively well the rotation and infalling motions, hence we vary the parameters within this range with 50\,au steps.
Together with the stellar mass, the radii values explored are constrained well enough that they can match the velocity distribution, hence here we determine the models that best match the shape of the streamer by visual inspection.
While a centrifugal radius of 500\,au fits the morphology of the streamer better (particularly the northernmost position of the $80\sigma$ contour level in Fig.~\ref{fig:continuum}B) than the previous value of 400\,au, equally good matches can be obtained with a centrifugal radius between 450 and 550\,au.
Same as in \cite{Olguin2023}, an initial polar angle of 80\degr\ reproduces the position of the streamers the best, while the best initial azimuthal angle is 55\degr\ with models with $\pm5$\degr\ angles matching the spine ($10\sigma$ level and higher) relatively well.
This result is plotted over the streamer in the first moment map in Fig.~\ref{fig:continuum}B  (outer streamline model) and the PV diagram in Fig.~\ref{fig:pvmap}.
For the red-shifted streamer, we did not perform any further optimization other than updating the centrifugal radius (see Fig.~\ref{fig:continuum}B).

\begin{table}[tbhp] 
        \centering
        \caption{\textbf{Streamline model parameter ranges per streamer.}
                The model parameters are: centrifugal radius ($r_c$), initial polar angle ($\theta_0$), initial azimuthal angle ($\phi_0$), and initial radial velocity ($v_{r0}$).
                The values in brackets indicate the parameter steps in the same units.}
        \label{tab:streamline:params} 

        \begin{tabular}{lcccc} 
                \\
                \hline
                Streamer & $r_c$ & $\theta_0$ & $\phi_0$ & $v_{r0}$\\
                         & (au)  & (\degr)    & (\degr)  & (\kms)\\
                \hline
                Outer blue & 400--600 (50) & 75--85 (5) & 50--70 (5)   & 0\\
                Inner blue & 200           &         89 & 140--195 (5) & 0, 0.1, 0.25, 0.5, 0.75\\
                \hline
        \end{tabular}
\end{table}

Additionally, we use the IRE model in FERIA \cite{Oya2022} to compare with the PV diagram in \cite{Olguin2023} with the updated centrifugal barrier (250\,au).
Fig.~\ref{fig:supp:model} shows a good agreement between the model and observation.
Note that in \cite{Olguin2023} the blue-shifted outflow defining the rotation axis in FERIA standard was assumed to point southward, resulting in a envelope P.A. of 125\degr\ based on the line roughly passing through the three continuum peaks.
However, the blue-shifted outflow and the geometry of the streamer indicate that the rotation axis points northward.
We thus update the P.A. to --55\degr\ and adapt the rotation direction accordingly.

\subsubsection*{Inner blue streamer modeling}

The streamline model of \cite{Mendoza2009} describes the trajectory of a particle under a gravitational potential with an initial radial velocity to the center of the potential and angular momentum.
In order to describe the motions of the blue-shifted streamer, we assume an additional system inside the centrifugal radius (inner component in Fig.~\ref{fig:sketch}A) where the trajectory is in the mid-plane orthogonal to the rotation axis, i.e., the trajectory of the large scale streamer changes once it joins the mid-plane.
Note that this is an approximation to match the end of the outer blue streamer with the inner streamer.
In the inner system, the initial radius where the particle falls, $r_0$, is given by the centrifugal radius, while the final radius (which would correspond to a second centrifugal radius), $r_f$, is set to 200\,au (the inner radius in Fig.~\ref{fig:pvmap}).
We also tried a final radius of 60\,au (the radius of the central continuum source), but we could not find a model that matches the inner streamer shape.
Since the \cite{Mendoza2009} model is undefined at a polar angle of 90\degr, we set the initial polar angle to 89\degr\ and tried values of 85\degr\ and 88\degr\ in case there were models that do not converge at 89\degr.
We varied the initial azimuthal angle and initial radial velocity (Table~\ref{tab:streamline:params}).
After a couple of iterations the initial azimuthal angle range was limited to 140--155\degr.
An angle of 145\degr\ matches the end position of the outer blue streamer the best (Fig.~\ref{fig:continuum}B).
Models with 0\,\kms\ initial velocity do not converge at 89\degr, but values closer to 0\,\kms\ reproduce the shape of the inner blue streamer the best, hence we select a value of 0.1\,\kms.
A similar fit can be obtained by a model with 0\,\kms\ but initial polar angle of 88\degr.
The central mass, $M_c$, was set to 10\,\msun\ \cite{Olguin2023}.

Note that the final radius is related to the angular velocity, $\Omega$, initial radius, and central mass ($M_c$; in this case the mass of ALMA1) by:
\begin{equation}
    r_f = \frac{r_0^4  \Omega^2}{G M_c}
\end{equation}
with $G$ the gravitational constant.
Thus a larger angular velocity, like the one required to match the velocity distribution at the initial radius of the streamer in Fig.~\ref{fig:continuum}B and \ref{fig:pvmap}, would not match the extent of the streamer under the assumptions above.

\subsubsection*{Velocity distributions}

The velocity distribution for Keplerian rotation is given by an azimuthal velocity:
\begin{equation}\label{eq:keplerian}
    v_\phi = \sqrt{\frac{GM_c}{r}}~.
\end{equation}
\noindent We considered that the inner blue streamer is in the mid-plane, so the distance to ALMA1, $r$, is independent whether spherical or cylindrical coordinates are used.
For free-fall the radial velocity is given by \cite{Tang2012}:
\begin{equation}\label{eq:freefall}
    v_r = \sqrt{\frac{2GM_c}{r}}~.
\end{equation}

On the other hand the velocity distribution of the IRE model \cite{Oya2022} is given by an azimuthal velocity
\begin{equation}
    v_{\phi} = \frac{1}{r}\sqrt{2GM_c r_{cb}}
\end{equation}
and a radial velocity
\begin{equation}\label{eq:infall}
    v_r = \frac{1}{r}\sqrt{2GM_c (r- r_{cb})}
\end{equation}
where $r_{cb}$ corresponds to the centrifugal barrier radius.

Projection of the velocities along the line of sight where calculated following \cite{Tang2012}.
The deprojected radial distance to the source was calculated assuming that the mid-plane defined by the axes $(x, y)$ is inclined an angle $i$ with respect to the plane of the sky defined by the axes $(x', y')$.
For a point located at a distance $r'$ to the source in the plane of the sky, the position of a point can be described by $(r' \cos \phi, r' \sin \phi)$.
For $x=x'$ and $x$ along the major axis of the mid-plane, the azimutal angle can be expressed as $\phi= {\rm P.A.} - {\rm P.A}._{\rm disk}$, with P.A. the position angle of a given point and ${\rm P.A.}_{\rm disk}{=}-55$\degr\ (see above)  the position angle of the mid-plane.
We can then express the distance from the $y$-axis as
\begin{equation}
    y = \frac{y'}{\cos i}=r' \frac{\sin \phi}{\cos i}
\end{equation}
and thus the deprojected distance to ALMA1 as:
\begin{equation}
    r = \sqrt{x^2 + y^2} = r' \sqrt{\frac{\sin^2 \phi}{\cos^2 i} + \cos^2\phi}
\end{equation}
with the envelope inclination angle $i=65$\degr\ \cite{Olguin2023} in FERIA and \cite{Tang2012} standard (0\degr\ for face-on configuration).

\subsubsection*{Replenishing times}

We determine the times it will take the inner streamers to maintain a constant gas mass in ALMA1.
First we estimate the mass of the gas content around the central young star and the inner streamers from their continuum emission assuming optically thin dust:
\begin{equation}
    M_d = \frac{S_{\nu} d^2}{R_{dg} \kappa_{\nu} B_{\nu}(T)}
\end{equation}
where $d=3.1$\,kpc is the distance to the source, $R_{dg}=0.01$ is the dust-to-gas mass ratio, $\kappa_{\nu}=1$\,cm$^2$\,g$^{-1}$ is the dust opacity at 1.3\,mm \cite{Ossenkopf1994}, and $B_{\nu}(T)$ the Planck blackbody function.
The flux densities, $S_\nu$, of ALMA1 is 7.7\,mJy (from Gaussian fit, see above), the blue-shifted inner streamer (measured roughly around the 20$\sigma$ continuum level within the 500\,au radius) is 20.7\,mJy and the red-shifted streamer is 11.1\,mJy (measured roughly around the 10$\sigma$ continuum level; all measured in the primary beam corrected image).
The region files used for the calculation of the inner streamer masses are publicly available online (see Data and materials availability).
The region files used for the calculation of the inner streamer masses are publicly available online (see Data and materials availability)
We use a single temperature of 100\,K for the central source and streamers, which corresponds roughly to the dust brightness temperature.
In the optically thick regime the gas temperature would be equal to the brightness temperature.
This temperature is slightly higher than the average temperature of the outer streamers found by \cite{Olguin2023} (93\,K for the blue shifted streamer). 
Note that the gas in ALMA1 is likely hotter than the streamers, making its mass an upper limit under the optically thin approximation.
The resulting masses are 0.24\,\msun\ for the gas in ALMA1, 0.6\,\msun\ for the inner blue streamer and 0.3\,\msun\ for the red-shifted streamer.

We measure a fastest possible rate by assuming free-fall, and a lower-bound infall rate assuming viscous dissipation in a disk.
The free-fall time is given by:
\begin{equation}
    t_{ff} = \sqrt{\frac{3 \pi}{ 32 G \rho}}
\end{equation}
where $\rho=M / (\frac{4}{3} \,\pi R_c^3)$ is the density, in this case assumed to be a spherical distribution with radius $R_c=61$\,au (half the deconvolved size of the central source) and a total mass $M=10$\,\msun.
On the other hand, the viscous accretion time is given by \cite{Shakura1973}:
\begin{equation}
    t_{acc} \approx \frac{r^2 \Omega}{3\alpha c_s^2}
\end{equation}
with $\Omega = v_r / r$ the angular velocity, $\alpha$ the viscosity parameter, and $c_s = \sqrt{k_B T/ \mu m_H}$ the speed of sound where $k_B$ is the Boltzmann constant and $m_H$ is the mass of the hydrogen atom.
The velocity $v_r$ corresponds to the Keplerian velocity (Equation~\ref{eq:keplerian}).
For marginally unstable protostellar disks during the main accretion phase and at high accretion rates, $\alpha$ ranges between 0.1 and 1 \cite{Kuiper2011,Yamamuro2023}.
The molecular weight per gas particle is $\mu=2.33$ \cite{Kauffmann2008}.
For the central region, the free-fall time is ${\sim}25$\,yr and the viscous time range is ${\sim}10^3-10^4$\,yr for $r=R_c$ and $\alpha=1$ and 0.1, respectively.
Similarly, the viscous accretion time is ${\sim}10^4-10^5$\,yr at the centrifugal radius (500\,au).

To estimate the infall rates for the free-fall time, we first assume a constant free-fall velocity, $v_{ff}$, given by Equation~\ref{eq:freefall} of 6\,km\,s$^{-1}$ for a central mass of 10\,\msun\ and a radius of $r=500$\,au.
The infall rate is given by \cite{Kirk2013}:
\begin{equation}
    \dot{M} = \frac{v_{ff} M_d}{l}
\end{equation}
where $l$ is the length of the inner streamer and assumed to be 500\,au in this case.
This gives infall rates of $1.6\times10^{-3}$\,\msun\,yr$^{-1}$ and $0.9\times10^{-3}$\,\msun\,yr$^{-1}$ for the inner blue and red-shifted streamers, respectively.
These are roughly one order of magnitude higher than the infall rate from the outer streamers estimated by \cite{Olguin2023} using their free-fall time and other methods including the velocity distribution (${\sim}10^{-4}$\,\msun\,yr$^{-1}$). 
For the viscous accretion timescale, the infall rate ranges between $M_d/t_{acc}{\sim}10^{-6}-10^{-5}$\,\msun\,yr$^{-1}$ for each streamer.
Therefore, it would take the blue-shifted streamer between ${\sim}150$\,yr (free-fall) and ${\sim}10^3-10^4$\,yr (viscous) to accumulate the same amount of gas in the putative disk in ALMA1.
Note that the latter assumes that the streamers are in a disk, which is likely not the case.
Similarly, it would take the red-shifted streamer between ${\sim}280$\,yr to $10^4-10^5$\,yr to replenish the gas mass in ALMA1, while together it would take ${\sim}10^2-10^4$\,yr to do the same.

To complement these calculations, we make an order-of-magnitude estimate of the forces exerted by radiation and the inflow.
Following \cite{Keto2006}, the force exerted by the young star radiation field assuming spherical symmetry and absorption of all the stellar radiation by the dust is of the order of $L/c$ with $L$  the luminosity and $c$  the speed of light, while the force from the inflow under spherical accretion is of the order of $\dot{M}v_{in}$ with $v_{in}$ the velocity of the flow.
Assuming that the luminosity of the clump $L=2.5\times10^{4}$\,\lsun\ \cite{Urquhart2018} is dominated by G336 ALMA1, the force exerted by the source radiation field is of the order of $10^{22}$\,N.
On the other hand, the velocity of the flow at the impact position (radius of 61\,au) is ${\sim}17$\,\kms\ from Equation~\ref{eq:freefall}, thus the force exerted by the inflow resulting from the blue-shifted streamer is of the order of $10^{24}$\,N.


\clearpage 

%

%
%
%
%
%
%


\section*{Acknowledgments}

The authors would like to thank Dr. James Jackson for contributing in improving the flow of the text, and the thoughtful comments from the anonymous reviewers.

\paragraph*{Funding:}
F.O. and Y.O. acknowledges the support of the NAOJ ALMA Joint Scientific Research Program grant No. 2024-27B.
F.O. and H.-R.V.C. acknowledge the support from the National Science and Technology Council (NSTC) of Taiwan grants NSTC 112-2112-M-007-041 and NSTC 112-2811-M-007-048. 
P.S. was partially supported by a Grant-in-Aid for Scientific Research (KAKENHI Number JP22H01271 and JP23H01221) of JSPS. 
A.G. acknowledges support from the NSF via grants AAG 2008101 and CAREER 2142300.
X.L. acknowledges support from the National Key R\&D Program of China (No. 2022YFA1603101), the Strategic Priority Research Program of the Chinese Academy of Sciences (CAS) Grant No. XDB0800300, the National Natural Science Foundation of China (NSFC) through grant Nos. 12273090 and 12322305, the Natural Science Foundation of Shanghai (No. 23ZR1482100), and the CAS “Light of West China” Program No. xbzg-zdsys-202212.
Y.O. acknowledges the support by Grant-in-Aids from Ministry of Education, Culture, Sports, Science, and Technologies of Japan (KAKENHI; 21K13954, 25K07367).
\paragraph*{Author contributions:}
Conceptualization: PS, FO, AG, HRVC, XL, FN, QZ, AEG\\
Methodology: FO, PS, HRVC, KEIT, FN, QZ, QL, AEG\\
Software: FO, AG, YO\\
Validation: FO, PS, AG, SL, AEG \\
Formal analysis: FO, PS, KEIT, HRVC, XL, QL, AEG\\
Investigation: FO, PS, AG, XL, FN, QZ\\
Resources: HRVC, PS\\
Data curation: FO, PS\\
Writing – original draft: FO, PS, HRVC, FN\\
Writing – review \& editing: FO, PS, AG, HRVC, KEIT, XL, KM, SL, YC, QZ, QL, YO, TS, MS, AEG\\
Visualization: FO\\
Supervision: PS, HRVC\\
Project administration: PS\\
Funding acquisition: PS, HRVC, YO, XL
\paragraph*{Competing interests:}
The authors declare they have no competing interest.
\paragraph*{Data and materials availability:} All data needed to evaluate the conclusions in the paper are present in the paper and/or the Supplementary Materials.
This paper makes use of the following ALMA data: ADS/JAO.ALMA\#2017.1.00237.S and ADS/JAO.ALMA\#2022.1.00700.S ALMA is a partnership of ESO (representing its member states), NSF (USA) and NINS (Japan), together with NRC (Canada), MOST and ASIAA (Taiwan), and KASI (Republic of Korea), in cooperation with the Republic of Chile. The Joint ALMA Observatory is operated by ESO, AUI/NRAO and NAOJ.
The original observational data are available through the ALMA archive located at \url{https://almascience.nao.ac.jp/aq/} with the program codes listed above.
The delivered data was reduced and further processed with \textsc{CASA} \cite{2022PASP..134k4501C}, with their calibration scripts available through the ALMA archive.
The underlying maps and models needed to reproduce the figures and results, and data cube cutouts are publicly available as Zenodo 10.5281/zenodo.15354559 (\url{https://zenodo.org/records/15354559}), while the scripts and auxiliary files necessary to reproduce these data, results and figures are publicly available as Zenodo 10.5281/zenodo.15362023 (\url{https://zenodo.org/records/15362023}).
The data necessary to produce fig.~\ref{fig:supp:model} is available in \cite{Olguin2023}.
Continuum and continuum-subtracted visibilities were produced with \textsc{GoContinuum} (v3.0.0 \cite{fernando_2025_15314190}).
The data cubes were produced using \textsc{YCLEAN} \cite{Contreras18,2018zndo...1216881C}.
Streamline models were calculated using the Python package \textsc{velocity\_tools} (v1.1) available at \url{https://github.com/jpinedaf/velocity_tools}, while IRE models were calculated using \textsc{FERIA} \cite{Oya2022} available at \url{https://github.com/YokoOya/FERIA} (\url{https://doi.org/10.1088/1538-3873/ac8839}).
Position-velocity diagrams were calculated using \textsc{pvextractor} available at \url{https://pvextractor.readthedocs.io/en/latest/index.html}.
This work made use of \textsc{Astropy} (\url{http://www.astropy.org}): a community-developed core Python package and an ecosystem of tools and resources for astronomy \cite{astropy:2013, astropy:2018, astropy:2022}.
The Scientific color maps vik and lipari \cite{crameri_2023_8409685} are used in this study to prevent visual distortion of the data and exclusion of readers with color-vision deficiencies \cite{2020NatCo..11.5444C}.

\paragraph{List of Supplementary Materials:}
Figures \ref{fig:supp:mom0}--\ref{fig:supp:sov1}
\clearpage
\newpage


\renewcommand{\thefigure}{S\arabic{figure}}
\renewcommand{\thetable}{S\arabic{table}}
\renewcommand{\theequation}{S\arabic{equation}}
\renewcommand{\thepage}{S\arabic{page}}
\setcounter{figure}{0}
\setcounter{table}{0}
\setcounter{equation}{0}
\setcounter{page}{1} 


\begin{center}
\section*{Supplementary Materials for\\ \scititle}
    Fernando~A.~Olguin$^{\ast}$,
    Patricio~Sanhueza,
    Adam~Ginsburg, 
    Huei-Ru~Vivien~Chen,
    Kei~E.~I.~Tanaka,
    Xing~Lu,
    Kaho~Morii,
    Fumitaka~Nakamura,
    Shanghuo~Li,
    Yu~Cheng,
    Qizhou~Zhang, 
    Qiuyi~Luo,
    Yoko~Oya,
    Takeshi~Sakai,
    Masao~Saito,
    Andr\'es~E.~Guzm\'an \\
	\small$^\ast$Corresponding author. Email: f.olguin@yukawa.kyoto-u.ac.jp
\end{center}

\subsubsection*{This PDF file includes:}
\begin{figure}
\centering
\includegraphics[width=0.8\textwidth]{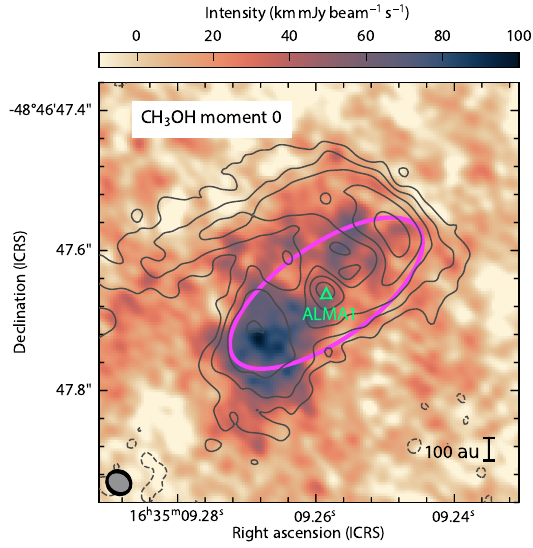}
\caption{\textbf{Zeroth order moment of CH$_3$OH in color scale with continuum in gray contours.}
The contour levels are the same as in Fig.~\ref{fig:continuum}. 
The pink ellipse shows the projected size of a disk of radius 500\,au.
The synthesized beam are shown in the bottom left corner for the zeroth order moment map (black) and continuum (gray).
}\label{fig:supp:mom0}
\end{figure}

\begin{figure}
\centering
\includegraphics[width=0.8\textwidth]{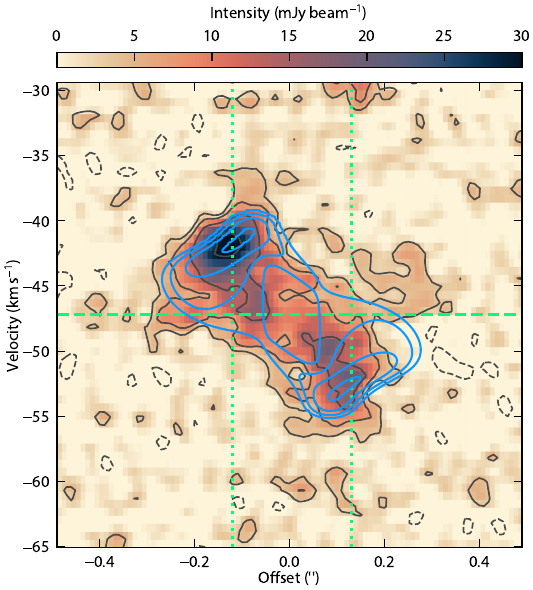}
\caption{\textbf{Observed position-velocity diagram of CH$_3$OH along the rotation direction (${\bf {\rm \bf P.A.}=-55}$\degr) from \cite{Olguin2023} in color scale and gray contours, and updated IRE model with a centrifugal barrier of 250\,au in blue contours.}
Gray contour levels are $-6$, $-3$, 3, 6, 12 and $24\times\sigma$ with $\sigma=1.2$\,mJy\,beam$^{-1}$, while blue contours are at the same levels after re-scaling to match the CH$_3$OH peak emission.
The horizontal green dashed line corresponds to the systemic velocity ($-47.2$\,\kms), and the vertical dotted lines correspond to the offsets of the continuum peaks.
}\label{fig:supp:model}
\end{figure}

\begin{figure}
\centering
\includegraphics[width=0.68\textwidth]{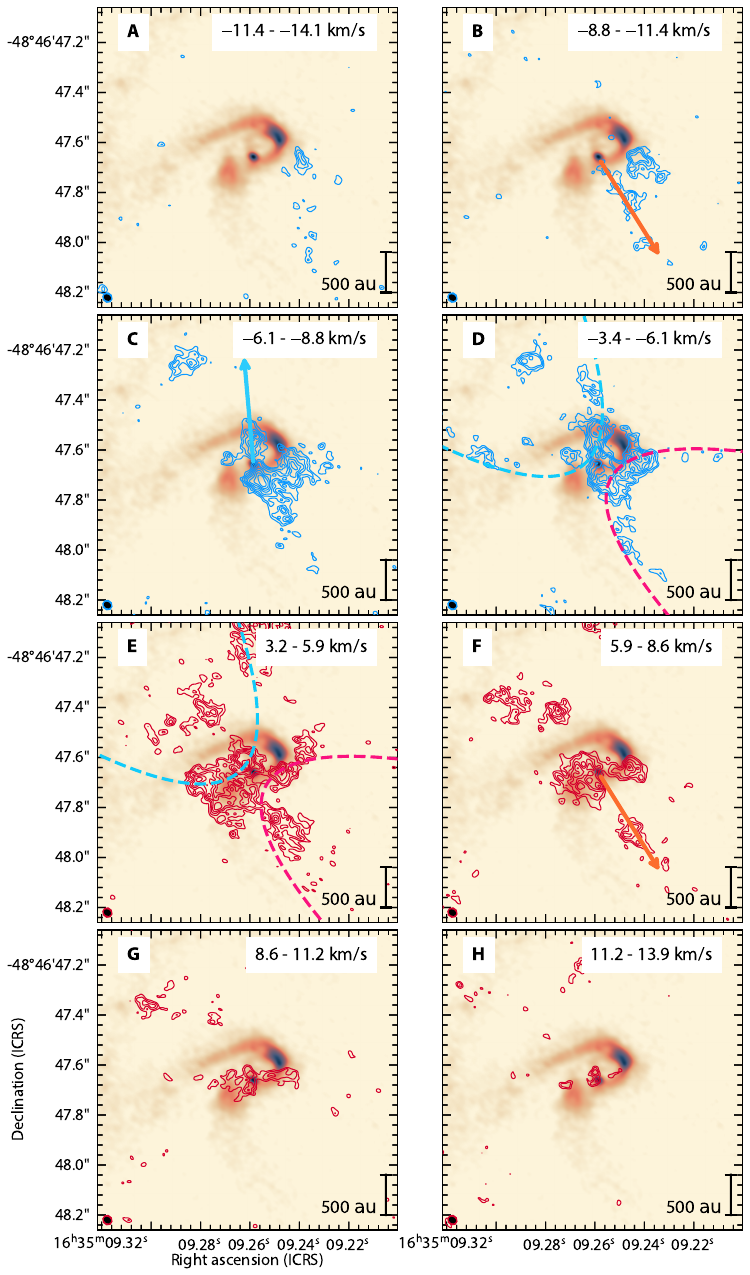}
\caption{\textbf{SO $\bf ^3\Sigma\,v{=}0\,J_K=6_5-5_4$ channel averages in contours over 1.3\,mm continuum.}
The contour levels are 3, 4, 5, ... $\times \sigma$ with $\sigma=2.5$\,mJy\,beam$^{-1}$\,km\,s$^{-1}$.
The blue and red dashed parabolas indicate the position and direction of the respective outflow cavities.
The arrows indicate the direction of shocked gas likely tracing a jet component.
The synthesized beams are shown in the bottom left corner for the  continuum (black) and contours (respective color).
}\label{fig:supp:sov0}
\end{figure}

\begin{figure}
\centering
\includegraphics[width=0.68\textwidth]{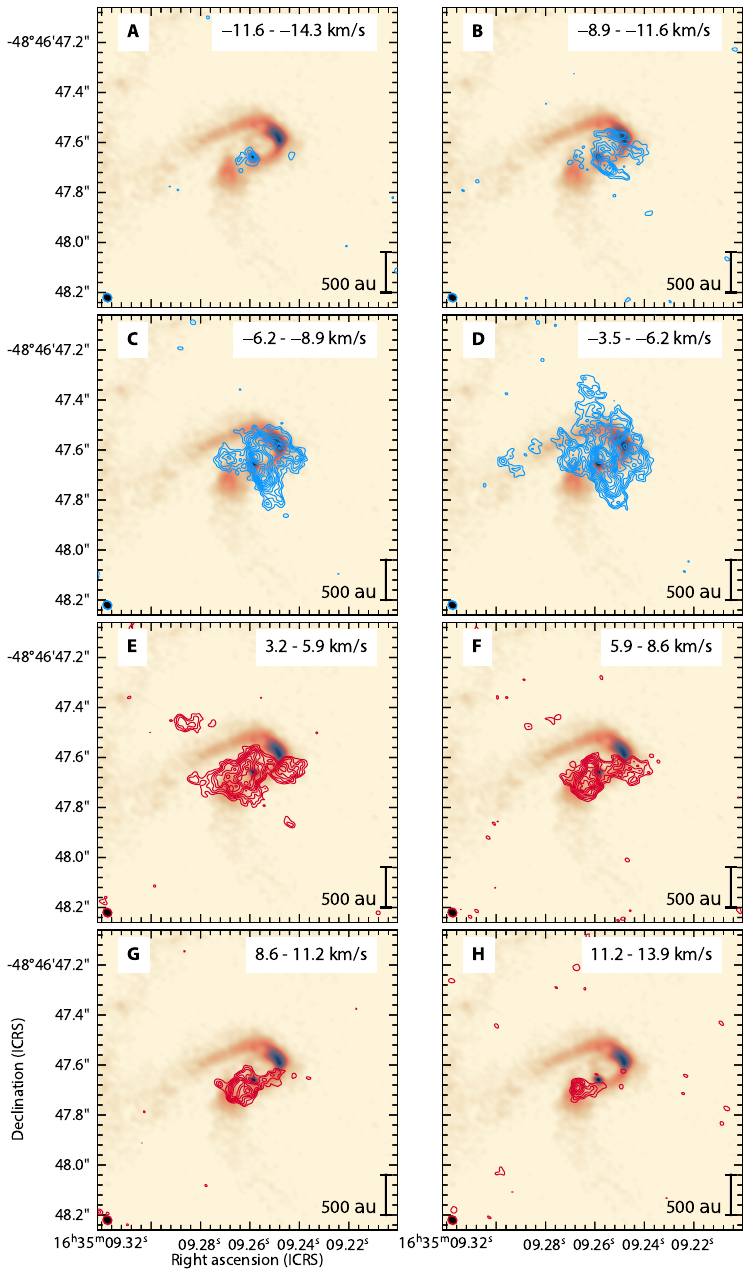}
\caption{\textbf{SO $\bf ^3\Sigma\,v{=}1\,J_K=6_5-5_4$ channel averages in contours over 1.3\,mm continuum.}
The contour levels are 3, 4, 5, ... $\times \sigma$ with $\sigma=2.8$\,mJy\,beam$^{-1}$\,km\,s$^{-1}$.
The synthesized beams are shown in the bottom left corner for the  continuum (black) and contours (respective color).
}\label{fig:supp:sov1}
\end{figure}

Supplementary Figs.~\ref{fig:supp:mom0} to \ref{fig:supp:sov1}.

Fig.~\ref{fig:supp:mom0} shows the zeroth order moment map of the CH$_3$OH $J_{K_a,K_c} =18_{3,15}-17_{4,14}\,A,\,v_t=0$ transition.
Fig.~\ref{fig:supp:model} shows the position-velocity diagram along a cut passing roughly through the three continuum peaks (${\rm P.A.} = -55$\degr).
Figs.~\ref{fig:supp:sov0} and \ref{fig:supp:sov1} show averaged channel maps of SO $^3\Sigma\,v=0\,J_K=6_5-5_4$ and $^3\Sigma\,v=1\,J_K=6_5-5_4$ transitions, respectively.

\end{document}